\documentclass[10pt,a4paper]{article}

\usepackage[english]{babel} 

\usepackage[T1]{fontenc}
\usepackage{tgheros}

\usepackage[top=2.5cm,bottom=2.5cm,left=2.5cm,right=2.5cm,columnsep=1.2cm]{geometry} 
\usepackage{multicol} 
\usepackage{graphicx} 
\usepackage{titlesec} 
\usepackage{caption} 
\usepackage{enumitem} 
\usepackage{parskip} 
\usepackage{amsmath} 
\usepackage{microtype} 
\usepackage{fancyhdr} 
\usepackage{float} 
\usepackage{footmisc} 
\usepackage{cite} 
\usepackage{tabularx}

\usepackage{hyperref}
\hypersetup{hidelinks}
\usepackage{booktabs}
\usepackage{multirow}
\usepackage{amssymb}
\usepackage{tikz}
\usetikzlibrary{shapes,arrows, positioning,arrows.meta}
\usepackage{bm}
\titleformat{\section}{\bfseries\normalsize}{\thesection.}{1em}{}
\titleformat{\subsection}{\bfseries\normalsize}{\thesubsection.}{1em}{}
\titleformat{\subsubsection}{\itshape\normalsize}{\thesubsubsection.}{1em}{} 
\DeclareCaptionFormat{customhang}{
 \textit{#1#2}%
 \hspace{0.25cm}%
 \textit{#3}
}
 \usetikzlibrary{positioning,matrix,arrows.meta,decorations.pathreplacing,calc}
\tikzset{
 arr/.style={-Stealth,thick},
 darr/.style={-Stealth,thick,dotted}, 
 ddarr/.style={-Stealth,thick,dash pattern=on 2pt off 2pt on 6pt off 2pt} 
}
\tikzset{
 lab/.style={midway,sloped,fill=white,inner sep=0.8pt,outer sep=0pt}, 
 labalt/.style={midway,sloped,fill=white,inner sep=0.8pt,outer sep=0pt,xshift=8mm}, 
}

\definecolor{slackcolor}{HTML}{B00020} 
\definecolor{pvcolor}{HTML}{1F6EA3} 
\definecolor{pqcolor}{HTML}{2E8B57} 

\setlist[itemize]{left=0.63cm, label=\textbullet, itemsep=0pt, topsep=6pt}

\title{\textbf{\fontsize{14pt}{14pt}\selectfont Impact of Training Dataset Size for ML Load Flow Surrogates}}
\author{
\fontsize{12pt}{12pt}\selectfont\textbf{Timon Conrad$^{1*}$, Johann Jäger$^{3}$} \\
\fontsize{10pt}{10pt}\selectfont\textit{Institute of Electrical Energy Systems, Friedrich-Alexander-Universität Erlangen-Nürnberg, Erlangen} \\
\vspace{8pt} \\
\fontsize{12pt}{12pt}\selectfont\textbf{Changhun Kim$^{2}$, Andreas Maier$^{4}$, Siming Bayer$^{5}$} \\
\fontsize{10pt}{10pt}\selectfont\textit{Pattern Recognition Lab, Friedrich-Alexander-Universität Erlangen-Nürnberg, Erlangen}
}
\date{}

\begin{document}

\twocolumn[
\vspace*{-1.5em}
\begin{center}
 {\fontsize{9pt}{11pt}\selectfont Oberlausitzer Energiesymposium 2025 \& Zittauer Energieseminar, Zittau, Deutschland, 25./26. November 2025}
\end{center}
\vspace{0.5em}

\maketitle
\vspace{-1em}
{\fontsize{10pt}{12pt}\selectfont
Efficient and accurate load flow calculations are a bedrock of modern power system operation. Classical numerical methods such as the Newton–Raphson algorithm provide highly precise results but are computationally demanding, which limits their applicability in large-scale scenario studies and optimization with time-critical contexts. Research has shown that machine learning approaches can approximate load flow results with high accuracy while substantially reducing computation time.
Sample efficiency, i.e. their ability to achieve high accuracy with limited training dataset size, is still insufficiently researched, especially in grids with a fixed topology. This paper considers a systematic investigation of the sample efficiency of a Multilayer Perceptron and two Graph Neural Networks variants on a dataset based on a modified IEEE 5-bus system. The results for this grid size show that Graph Neural Networks achieve the lowest losses. However, the availability of large training datasets remains the dominant factor for performance over architecture. Code: \url{https://github.com/timonOconrad/loadflow-ai}
}
\vspace{1em}
]

\section{Introduction}

The AC load flow calculation is widely used as a fundamental component of modern power system operation, particularly in the context of congestion identification and resolution under frameworks such as Redispatch~2.0. The Newton-Raphson (N-R) algorithm, a classical numerical method, has been extensively employed in the solution of non-linear equations of AC load flow. This method is known for its high accuracy, good convergence rate and robustness. However, its computational cost can become prohibitive in large-scale scenario studies or optimization in contexts requiring rapid responses in large-scale grids. \cite{kundur1994power}\\
Surrogate models based on Multilayer Perceptrons (MLPs) have been explored for several decades. An early contribution was made by \cite{nguyen1995neural}, who proposed a neural network architecture that emulates the Newton--Raphson method to estimate voltage magnitudes and phase angles. This was followed by a number of further developments that applied MLPs to various load flow applications, e.g. \cite{PaucarRider2002, PhamLi2022NNPowerFlow}.\\
More recently, Graph Neural Networks (GNNs) have been proposed as a more structure-aware alternative to MLPs. GNNs exploit the grid topology explicitly, modelling buses as nodes and electrical connections as edges. By embedding the power system as a graph, GNNs are capable of leveraging local and global structural information through message passing schemes. This architectural property enables improved generalization across varying power injections, particularly in scenarios where the grid topology is fixed.\\
Several studies have reported that GNN-based surrogates outperform classical MLPs (mentioned as fully connected networks) in accuracy and training time for small \cite{lin2023powerflownet} and big \cite{lopez2022typed, donon2019graph} grids.\\
In \cite{donon2019graph}, a GNN-based architecture generalized to grids of different sizes (10–110 buses) despite being trained only on 30-bus systems. This transferability was not achievable with MLPs, which failed under varying grid structures due to their fixed input vectors. Furthermore, the proposed GNN converged faster in terms of iterations and achieved a twofold reduction in inference time compared to a commercial load flow solver.\\
Despite promising results in the application of GNNs and MLPs as surrogate models for load flow approximation, little attention has been paid to their sample efficiency, especially in fixed-topology settings. While previous work has focused on architectural innovations and generalization across topologies or grid sizes \cite{donon2019graph, lin2023powerflownet, lopez2022typed}, the relationship between training dataset size and model performance remains insufficiently understood.\\
This paper addresses this gap by presenting a comparison of three neural network architectures on a fixed topology. All experiments are conducted on a modified IEEE 5-bus system using the same inputs to ensure comparability across models.
\newpage The results provide a basis for quantifying sample efficiency in constrained scenarios and using these results to support the development of GNN-based surrogate models when scaling up to larger power grids.

\section{Dataset}
\label{sec:dataset}

The dataset\footnote{\url{https://github.com/timonOconrad/static-voltage-stability-AI}} used in this work consists of synthetically generated load flow cases based on a modified IEEE 5-bus system. Each entry contains the full set of load flow results, including voltages in real and imaginary part, active and reactive powers at each bus. The dataset comprises a total of 789{,}000 individual cases, stored in Parquet format for efficient processing.\\
The following section provides a brief overview of the data generation process, including the grid model and parameter variation. Details can be found in the associated thesis~\cite{conrad2023master}.

\subsection{Grid Model}
\label{sec:grid}
\begin{figure}[!ht]
\centering
\begin{tikzpicture}[busnode/.style={circle,draw,minimum size=10mm,inner sep=0pt,font=\bfseries}]

\node[busnode,fill=slackcolor,text=white,label=above:Bus 1] (B1) at (0,0) {S};
\node[busnode,fill=pvcolor,text=white,label=below:Bus 2] (B2) at (0,-3) {PV};
\node[busnode,fill=pqcolor,text=white,label=above:Bus 3] (B3) at (3,0) {PQ};
\node[busnode,fill=pqcolor,text=white,label=above:Bus 4] (B4) at (6,0) {PQ};
\node[busnode,fill=pqcolor,text=white,label=below:Bus 5] (B5) at (6,-3) {PQ};

\draw (B1) -- (B2) node[midway,left]{1--2};
\draw (B1) -- (B3) node[midway,above]{1--3};
\draw (B2) -- (B3) node[midway,left]{2--3};
\draw (B2) -- (B4) node[midway,below]{2--4};
\draw (B2) -- (B5) node[midway,below]{2--5};
\draw (B3) -- (B4) node[midway,above]{3--4};
\draw (B4) -- (B5) node[midway,right]{4--5};

\end{tikzpicture}

\caption{Modified IEEE 5-bus system used in data generation. S = Slack bus, PV = Generator bus, PQ = Load bus.}
\label{fig:grid}
\end{figure}
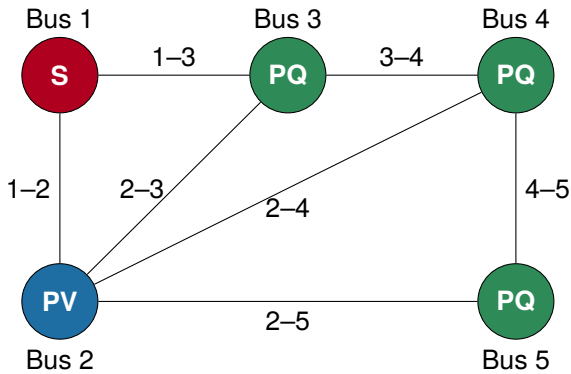

The grid used in this paper is based on the IEEE 5-bus system, with all line parameters and bus configurations identical to those described in the original specification ~\cite{bhandakkar2018ieee}. The only modification is that no load is connected to the PV bus (Bus 2), which distinguishes this model from the standard configuration. \\
The grid topology is illustrated in \autoref{fig:grid}.\\

\subsection{Data Generation}

The data was generated using AC load flow simulations performed in DIgSILENT PowerFactory. For each case, selected input parameters were randomly varied within physically meaningful limits. Specifically:

\begin{itemize}
 \item Active power at the PV bus (Bus 2) was varied between 0 and 199 MW.
 \item Active power demand at the PQ buses (Bus 3–5) was varied between 0 and 99 MW.
 \item Reactive power demand at the PQ buses (Bus 3–5) was varied between 0 and 99 MVAr.
\end{itemize}
All power values were generated using random sampling and rounded to integer values to simplify post-processing. The slack bus served as a fixed voltage reference and was not modified during the simulations.

\section{Architectures}
\label{sec:approaches}
\subsection{Graph Neural Network (GNN)}

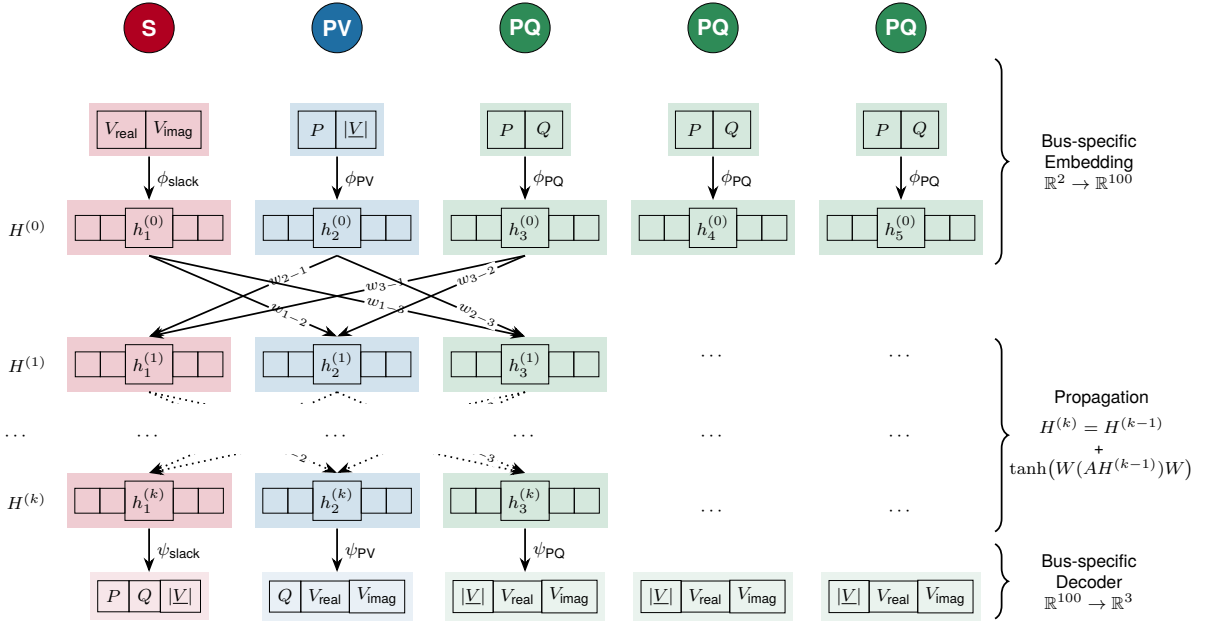
\begin{figure*}[t]
 \centering
\resizebox{\linewidth}{!}{
\begin{tikzpicture}[
 font=\footnotesize,
 bus/.style={circle,draw,minimum size=8mm,inner sep=0pt,font=\bfseries},
 features/.style={
 matrix of nodes,
 nodes={draw,minimum size=6mm,anchor=center},
 column sep=-\pgflinewidth, row sep=-\pgflinewidth,
 nodes in empty cells,
 ampersand replacement=\&
 },
 vector/.style={
 matrix of nodes,
 nodes={draw,minimum size=4mm,anchor=center},
 column sep=-\pgflinewidth, row sep=-\pgflinewidth,
 nodes in empty cells,
 ampersand replacement=\&
 },
 output/.style={
 matrix of nodes,
 nodes={draw,minimum size=5mm,anchor=center},
 column sep=-\pgflinewidth, row sep=-\pgflinewidth,
 nodes in empty cells,
 ampersand replacement=\&
 },
 arr/.style={-Stealth,thick},
 lab/.style={midway,sloped,fill=white,inner sep=0.8pt,outer sep=0pt}
]

\node[bus,fill=slackcolor,text=white] (B1) at (0,0) {S};
\matrix (F1) [features,fill=slackcolor!20,below=8mm of B1] {
 $ V_{\text{real}}$ \& $ V_{\text{imag}}$ \\
};

\node[bus,fill=pvcolor,text=white] (B2) at (3,0) {PV};
\matrix (F2) [features,fill=pvcolor!20,below=8mm of B2] {
 $P$ \& $|\underline{V}|$ \\
};

\node[bus,fill=pqcolor,text=white] (B3) at (6,0) {PQ};
\matrix (F3) [features,fill=pqcolor!20,below=8mm of B3] {
 $P$ \& $Q$ \\
};

\node[bus,fill=pqcolor,text=white] (B4) at (9,0) {PQ};
\matrix (F4) [features,fill=pqcolor!20,below=8mm of B4] {
 $P$ \& $Q$ \\
};

\node[bus,fill=pqcolor,text=white] (B5) at (12,0) {PQ};
\matrix (F5) [features,fill=pqcolor!20,below=8mm of B5] {
 $P$ \& $Q$ \\
};

\draw[decorate,decoration={brace,amplitude=6pt},thick]
 (13.5,-0.5) -- (13.5,-3.8)
 node[midway,xshift=15mm,align=center] {Bus-specific\\Embedding\\$\mathbb{R}^2 \to \mathbb{R}^{100}$};


\def\xstep{1}
\def\ystep{13mm}

\matrix (H0-1) [vector,fill=slackcolor!20,below=7mm of F1] { \& \& $h_1^{(0)}$ \& \& \\ };
\matrix (H0-2) [vector,fill=pvcolor!20,below=7mm of F2] { \& \& $h_2^{(0)}$ \& \& \\ };
\matrix (H0-3) [vector,fill=pqcolor!20,below=7mm of F3] { \& \& $h_3^{(0)}$ \& \& \\ };
\matrix (H0-4) [vector,fill=pqcolor!20,below=7mm of F4] {
 \& \& $h_4^{(0)}$ \& \& \\
};
\matrix (H0-5) [vector,fill=pqcolor!20,below=7mm of F5] {
 \& \& $h_5^{(0)}$ \& \& \\
};

\draw[arr] (F5.south) -- node[right] {$\phi_{\text{PQ}}$} (H0-5.north);
\draw[arr] (F4.south) -- node[right] {$\phi_{\text{PQ}}$} (H0-4.north);

\node[left=2mm of H0-1.west,anchor=east] {${H}^{(0)}$};
\draw[arr] (F1.south) -- node[right] {$\phi_{\text{slack}}$} (H0-1.north);

\draw[arr] (F2.south) -- node[right] {$\phi_{\text{PV}}$} (H0-2.north);

\draw[arr] (F3.south) -- node[right] {$\phi_{\text{PQ}}$} (H0-3.north);

\matrix (Hk-1) [vector,fill=slackcolor!20,below=\ystep of H0-1] { \& \& $h_1^{(1)}$ \& \& \\ };
\matrix (Hk-2) [vector,fill=pvcolor!20,below=\ystep of H0-2] { \& \& $h_2^{(1)}$ \& \& \\ };
\matrix (Hk-3) [vector,fill=pqcolor!20,below=\ystep of H0-3] { \& \& $h_3^{(1)}$ \& \& \\ };
\node at ($(H0-4.north)+(0,-2.5)$) {$\cdots$};
\node at ($(H0-5.north)+(0,-2.5)$) {$\cdots$};
\node[left=2mm of Hk-1.west,anchor=east] {${H}^{(1)}$};

\matrix (Hkp1-1) [vector,fill=slackcolor!20,below=\ystep of Hk-1] { \& \& $h_1^{(k)}$ \& \& \\ };
\matrix (Hkp1-2) [vector,fill=pvcolor!20,below=\ystep of Hk-2] { \& \& $h_2^{(k)}$ \& \& \\ };
\matrix (Hkp1-3) [vector,fill=pqcolor!20,below=\ystep of Hk-3] { \& \& $h_3^{(k)}$ \& \& \\ };
\node at ($(H0-4.north)+(0,-5)$) {$\cdots$};
\node at ($(H0-5.north)+(0,-5)$) {$\cdots$};
\node[left=2mm of Hkp1-1.west,anchor=east] {${H}^{(k)}$};

\newcommand{\connectrows}[3][arr]{%
 \draw[#1] (H#2-1.south) -- node[labalt] {\scriptsize $w_{1-2}$} (H#3-2.north);
 \draw[#1] (H#2-2.south) -- node[labalt] {\scriptsize $w_{2-1}$} (H#3-1.north);
 \draw[#1] (H#2-1.south) -- node[labalt] {\scriptsize $w_{1-3}$} (H#3-3.north);
 \draw[#1] (H#2-3.south) -- node[labalt] {\scriptsize $w_{3-1}$} (H#3-1.north);
 \draw[#1] (H#2-2.south) -- node[labalt] {\scriptsize $w_{2-3}$} (H#3-3.north);
 \draw[#1] (H#2-3.south) -- node[labalt] {\scriptsize $w_{3-2}$} (H#3-2.north);
}

\connectrows{0}{k} 
\connectrows[darr]{k}{kp1} 

\draw[decorate,decoration={brace,amplitude=6pt},thick]
 (13.5,-5) -- (13.5,-8.1)
 node[midway,xshift=17mm,align=center] {Propagation\\[1ex]
 ${H}^{(k)}={H}^{(k-1)}$\\+\;\\
 $\tanh\!\big({W}({A}{H}^{(k-1)}){W} \big )$};




\matrix (Out1) [output,fill=slackcolor!10,below=7mm of Hkp1-1] 
 { $P$ \& $Q$ \& $|\underline{V}|$ \\ };

\matrix (Out2) [output,fill=pvcolor!10,below=7mm of Hkp1-2] 
 { $Q$ \& $ V_{\text{real}}$ \& $ V_{\text{imag}}$ \\ };

\matrix (Out3) [output,fill=pqcolor!10,below=7mm of Hkp1-3] 
 { $|\underline{V}|$ \& $ V_{\text{real}}$ \& $ V_{\text{imag}}$ \\ };
\coordinate (yHk) at (Out3);

\matrix (Out4) [output,fill=pqcolor!10] at (H0-4 |- yHk)
 { $|\underline{V}|$ \& $ V_{\text{real}}$ \& $ V_{\text{imag}}$ \\ };
\matrix (Out5) [output,fill=pqcolor!10] at (H0-5 |- yHk)
 { $|\underline{V}|$ \& $ V_{\text{real}}$ \& $ V_{\text{imag}}$ \\ };

\draw[arr] (Hkp1-1.south) -- node[right] {$\psi_{\text{slack}}$}(Out1.north); 
\draw[arr] (Hkp1-2.south) -- node[right] {$\psi_{\text{PV}}$}(Out2.north);
\draw[arr] (Hkp1-3.south) -- node[right] {$\psi_{\text{PQ}}$}(Out3.north);

\draw[fill=white,draw=none] 
 ($(Hk-1.south west)+(0,-2mm)$) rectangle ($(Hk-3.south east)+(0,-10mm)$);
\node at ($(Hk-1.south)+(0,-7mm)$) {$\cdots$};
\node at ($(Hk-2.south)+(0,-7mm)$) {$\cdots$};
\node at ($(Hk-3.south)+(0,-7mm)$) {$\cdots$};
\node at ($(Hk-1.south west)+(-0.8,-7mm)$) {$\cdots$};

\coordinate (yHk) at ($(Hk-3.south east)+(0,-7mm)$);

\node at (H0-5 |- yHk) {$\cdots$};
\node at (H0-4 |- yHk) {$\cdots$};
\draw[decorate,decoration={brace,amplitude=6pt},thick]
 (13.5,-8.3) -- (13.5,-9.5)
 node[midway,xshift=15mm,align=center] {Bus-specific\\ Decoder\\$\mathbb{R}^{100} \to \mathbb{R}^{3}$};
 
\end{tikzpicture}
}
 \caption{Illustration of the architecture of the GNN Model with Bus-specific Decoder}
 \label{fig:gnn_full}
\end{figure*}

For the proposed GNN, the electrical grid $\mathcal{G} = (\mathcal{V}, \mathcal{E})$ is represented as an undirected graph with $|\mathcal{V}| = N$ buses (nodes) and transmission lines (edges). 
The architecture can be divided into 3 steps:
\begin{enumerate}
 \item Bus-specific Embedding
 \item Propagation (Message Passing)
 \item Decoding
\end{enumerate}

The architecture is illustrated in \autoref{fig:gnn_full} and described in more detail below.\\
\textbf{1. Bus-type-specific Embedding:} Each bus (node) is assigned to a type (slack, PV, PQ) based on the applied grid (\autoref{sec:grid}) and the feature vector ${f}_i \in \mathbb{R}^2$ \footnote{Note: In contrast to the calculation method used by Newton-Raphson, this implementation uses the real $\Re\{\bm{\underline{V}}\}$ and imaginary $\Im\{\bm{\underline{V}}\} $ part of the voltage, to avoid the usage of the voltage angle $\theta$ for reasons of normalization. In order to achieve the same vector size $\mathbb{R}^2$ for the slack bus type as for buses with PV or PQ type, the magnitude of the voltage $V =|\underline{V}|$ was omitted in the feature. As the bus-specific decoder for GNN1, also requires same vector size $\mathbb{R}^3$ for all bus types, the magnitude of the voltage $V =|\underline{V}|$ was inserted for the slack. For reasons of comparability, the output vector for GNN2 and the MLP also uses with these vectors.}
for each bus is constructed by the type (\ref{eq:bustype}).
\begin{equation}
\begin{aligned}
\text{Slack (Type 1):} & \quad {f}_i = \big[V_{\text{real}}, V_{\text{imag}}\big] \\
\text{PV (Type 2):} & \quad {f}_i = \big[P, |V|\big] \\
\text{PQ (Type 3):} & \quad {f}_i = \big[P, Q\big]
\end{aligned}
\label{eq:bustype}
\end{equation}\\
Each bus $i \in \mathcal{V}$ is first mapped into a $d$-dimensional embedding vector using a bus-specific function (\ref{eq:enbedding}).
\begin{equation}
{h}_i^{(0)} = \phi_{\text{type}(i)}({f}_i), 
\quad \phi: \mathbb{R}^2 \to \mathbb{R}^d.
\label{eq:enbedding}
\end{equation}\\
All bus embeddings results were collected in the node feature matrix (\ref{eq:featurematrix}) where each column corresponds to the latent representation of one bus.
\begin{equation}
{H}^{(0)} = \big[{h}_1^{(0)}, {h}_2^{(0)}, \dots, {h}_N^{(0)}\big]^\top
\in \mathbb{R}^{N \times d}.
\label{eq:featurematrix}
\end{equation}\\
\textbf{2. Propagation (Message Passing)} 
The connectivity used for message passing is encoded in the adjacency matrix (\ref{eq:adjustancmatrix}).
\begin{equation}
{A}_{ij} = 
\begin{cases} 
1 & \text{if } ({Y}_r)_{ij} \neq 0 \;\lor\; ({Y}_i)_{ij} \neq 0, \\
0 & \text{otherwise}.
\end{cases}
\label{eq:adjustancmatrix}
\end{equation}\\
${A} \in \{0,1\}^{N \times N}$, derived from the bus admittance matrix 
${Y}_{\text{bus}} = {Y}_r + j {Y}_i$.
Self-loops are removed by setting $A_{ii}=0$. Degree normalization is applied using the diagonal degree matrix ( \autoref{eq:adjustancmatrix2}).
${D}$ with $D_{ii} = \sum_j A_{ij}$:
\begin{equation}
{A} \;\leftarrow\; {D}^{-1}{A}.
\label{eq:adjustancmatrix2}
\end{equation}\\
Propagation is performed iteratively for $K$ steps. At each step, the node feature matrix is updated as in (\ref{eq:nodefeaturematrix}).
\begin{equation}
{H}^{(k)} 
= {H}^{(k-1)} 
 + 
 \tanh\!\big( ({A}\,{H}^{(k-1)})\,{W} \big)
\label{eq:nodefeaturematrix}
\end{equation}\\
The weight matrix ${W}$ with ${W} \in \mathbb{R}^{d \times d}$ is a trainable parameter 
that linearly transforms the aggregated neighbourhood information 
before the non-linearity is applied.  It ensures that messages received from neighbouring nodes 
are projected into the same latent space as the residual connection ${H}^{(k-1)}$. 
The residual connection stabilizes training and improves gradient flow.\\
\textbf{3. Decoding: } After $k$ propagation steps, the final node states 
$\{{h}_i^{(k)}\}_{i=1}^N$ are decoded into physical predictions. 
Two decoder strategies are used, which represent the difference in the two GNN architectures under consideration:
\begin{enumerate}[label=\textbf{\Alph*.}]
 \item \textbf{Global Decoder (GNN1):} 
 Aggregate all bus states by mean pooling (\ref{eq:meanpooling}) followed by a feed-forward mapping (\ref{eq:mapping}).
 \begin{equation}
 \bar{{h}} = \frac{1}{N} \sum_{i=1}^N {h}_i^{(K)},
 \label{eq:meanpooling}
 \end{equation}
 \begin{equation}
 {y} = \psi(\bar{{h}}), 
 \quad \psi: \mathbb{R}^d \to \mathbb{R}^{m}.
 \label{eq:mapping}
 \end{equation}
 This produces the same prediction vector ${y}$ as for the MLP.
 
 \item \textbf{Bus-specific Decoder (GNN2):} 
 Decode each bus separately using type-dependent decoding similar to the embedding step (\ref{eq:decoding2}).
 \begin{equation}
 y_i = \psi_{\text{type}(i)}(h_i^{(K)}),
 \quad \psi_{\text{type}(i)}: \mathbb{R}^d \to \mathbb{R}^{m_i}.
 \label{eq:decoding2}
 \end{equation}
 The full output is the concatenation of all buses (\ref{eq:pooling2}).
 \begin{equation}
 y = \big[y_1 \,\|\, y_2 \,\|\, \dots \,\|\, y_N\big].
 \label{eq:pooling2}
 \end{equation}
\end{enumerate}

\subsection{Multilayer Perceptron (MLP)}
As a baseline, a MLP is employed. 
To ensure comparability with the GNN, the same bus-specific features are concatenated into a single input vector (\ref{eq:vector}) which corresponds to the flattened feature (\ref{eq:bustype}) representation of all buses. \\ 
\begin{equation}
x \in \mathbb{R}^{10},
\label{eq:vector}
\end{equation}
The MLP predicts the same set of target variables (\ref{eq:targets}) as the GNN, representing the concatenated outputs across all buses.
\begin{equation}
{y} \in \mathbb{R}^{15},
\label{eq:targets} 
\end{equation}
Each hidden layer of the MLP applies a fully connected linear transformation 
followed by a non-linear activation (\ref{eq:activation}) where ${W}$ and ${b}$ denote the trainable weights and biases. 
\begin{equation}
{h}' = \tanh\!\left({W}{h} + {b}\right),
\label{eq:activation} 
\end{equation}
The input vector ${x}$ is thus successively transformed into higher-level representations, until the final output ${y}$ is obtained.\\
Unlike in the GNN, where message passing incorporates the grid topology via the adjacency matrix ${A}$, the MLP treats all input features as independent and fully connected. The weight matrices ${W}$ represent connections between all units of adjacent layers, without structural constraints from the power grid. This makes the MLP a topology-agnostic baseline against which the GNN can be compared.

\section{Experiments}

The experimental study was conducted using the three machine learning architectures introduced in \autoref{sec:approaches}. The MLP had one hidden layer with 64 neurons. The two GNNs had for $d=100$ and for $k= 5$. These parameters were determined initially and showed good results. 
The objective was to investigate how key hyperparameters influence model accuracy in the task of load flow approximation for all architectures. 
While the training dataset size constituted the primary focus, additional experiments were conducted to assess the effects of batch size and learning rate. The parameters relating to the architecture were not adjusted further to enable a comparison.
The explored hyperparameter configurations are summarized in \autoref{tab:hyperparams}.
\begin{table}[h]
\centering
\caption{Hyperparameter variations explored in the experiments.}
\label{tab:hyperparams}
\begin{tabular}{p{0.3\linewidth} p{0.6\linewidth}}
\toprule
Hyperparameter & Values varied \\
\midrule
Training size & 500, 1{.}000, 5{.}000, 10{.}000, 50{.}000, 100{.}000, 500{.}000 \\
Batch size & 16, 32, 64, 128 \\
Learning rate & $1\cdot10^{-4}$, $1\cdot10^{-3}$, $1\cdot10^{-2}$, $1\cdot10^{-1}$ \\
\bottomrule
\end{tabular}
\end{table}\\
All models were trained on the dataset described in \autoref{sec:dataset}. The considered cases were partitioned into 70\% training, 15\% validation, and 15\% testing splits. Features and targets were standardized using parameters derived from the training set. 
Training was performed for a maximum of 50 epochs using the Adam optimizer. The MLP was trained on flattened input vectors, whereas the GNNs operated directly on per-bus features and the fixed grid topology via the admittance matrix $Y$. Mean squared error (MSE) served as the primary loss function.
\begin{figure}[t]
 \centering
 \includegraphics[width=0.5\textwidth]{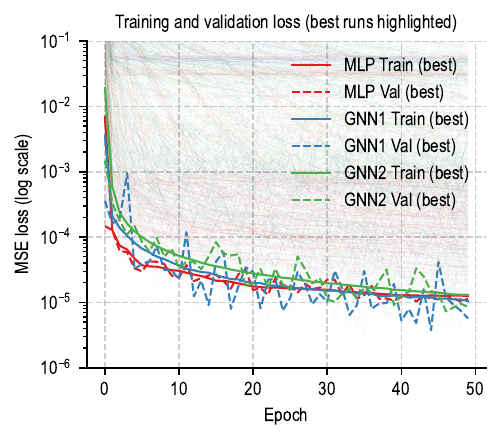}
 \caption{Training and Validation Loss Curves. The best configuration for each model is highlighted; all others are displayed with lower opacity.}
 \label{fig:loss_curves_allinone}
\end{figure}
\begin{table}[ht]
\centering
\caption{Top-3 runs per model ranked by lowest final validation loss.}
\label{tab:top_runs}
\begin{tabularx}{\linewidth}{l@{\hskip 2pt}c@{\hskip 2pt}c@{\hskip 2pt}c}
\toprule
 & MLP & GNN1 & GNN2 \\
\midrule
\multicolumn{4}{l}{\textbf{Run 1}} \\
Learning rate  & 0.001 & 0.001 & 0.0001 \\
Batch size & 32 & 32 & 32 \\
Cases & 500k & 500k & 500k \\
Train (MSE) & $1.25\cdot 10^{-5}$ & $1.05\cdot 10^{-5}$ & $1.31\cdot 10^{-5}$ \\
Val (MSE) & $1.14\cdot 10^{-5}$ & $5.66\cdot 10^{-6}$ & $8.55\cdot 10^{-6}$ \\
Test (MSE) & $1.14\cdot 10^{-5}$ & $5.65\cdot 10^{-6}$ & $8.51\cdot 10^{-6}$ \\
\midrule
\multicolumn{4}{l}{\textbf{Run 2}} \\
Learning rate & 0.001 & 0.001 & 0.0001 \\
Batch size & 16 & 16 & 16 \\
Cases & 500k & 500k & 500k \\
Train (MSE) & $1.47\cdot 10^{-5}$ & $1.06\cdot 10^{-5}$ & $8.92\cdot 10^{-6}$ \\
Val (MSE) & $1.18\cdot 10^{-5}$ & $6.49\cdot 10^{-6}$ & $9.17\cdot 10^{-6}$ \\
Test (MSE) & $1.18\cdot 10^{-5}$ & $6.51\cdot 10^{-6}$ & $9.15\cdot 10^{-6}$ \\
\midrule
\multicolumn{4}{l}{\textbf{Run 3}} \\
Learning rate & 0.001 & 0.0001 & 0.001 \\
Batch size & 64 & 32 & 16 \\
Cases & 500k & 500k & 500k \\
Train (MSE) & $1.87\cdot 10^{-5}$ & $9.31\cdot 10^{-6}$ & $2.04\cdot 10^{-5}$ \\
Val (MSE) & $2.00\cdot 10^{-5}$ & $7.36\cdot 10^{-6}$ & $1.28\cdot 10^{-5}$ \\
Test (MSE) & $2.01\cdot 10^{-5}$ & $7.36\cdot 10^{-6}$ & $1.28\cdot 10^{-5}$ \\
\bottomrule
\end{tabularx}
\label{tab:bestmodels}
\end{table}\\
As illustrated in~\autoref{fig:loss_curves_allinone} and ~\autoref{tab:bestmodels}, GNN1 (global decoder) achieved the lowest validation loss among all models, converging to $5.66 \cdot 10^{-6}$. 
However, its training dynamics were less stable, with noticeable fluctuations during early epochs. In contrast, the MLP exhibited a smoother and more stable convergence behavior, ultimately reaching a slightly higher but still competitive validation loss at $1.14 \cdot 10^{-5}$. It was surprising that the MLP still performed well compared to the GNNs, unlike \cite{lin2023powerflownet}, but this may be due to the fixed topology and probably no longer works as well with larger grids.
GNN2 (bus-specific decoder) positioned itself between these two extremes, with validation losses of $7.36 \cdot 10^{-6}$ but a less consistent progression. 
Overall, these results highlight that while GNN1 provides the best final validation accuracy, the MLP demonstrates the most robust and stable learning behavior across runs. 
The observed variance across runs indicates that additional hyperparameter optimization (e.g., learning rate schedules, depth, and regularization) will be required to fully exploit the potential of the GNN architectures.
\section{Results}
\begin{figure}[t!]
 \centering
 \includegraphics[width=0.5 \textwidth]{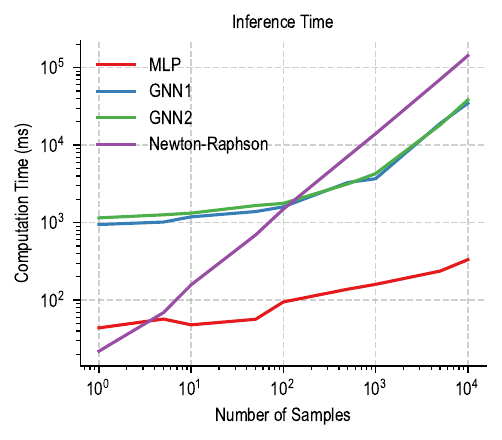}
 \caption{Inference time depending the number of samples on the best models (Run 1) or N-R algorithm}
 \label{fig:inftime}
\end{figure}

The final test using the test dataset ~\autoref{tab:bestmodels} shows comparable results to those of the validation dataset, with the GNN1 model proving to be the most effective architecture.
\begin{figure}[t!]
 \includegraphics[width=0.5 \textwidth]{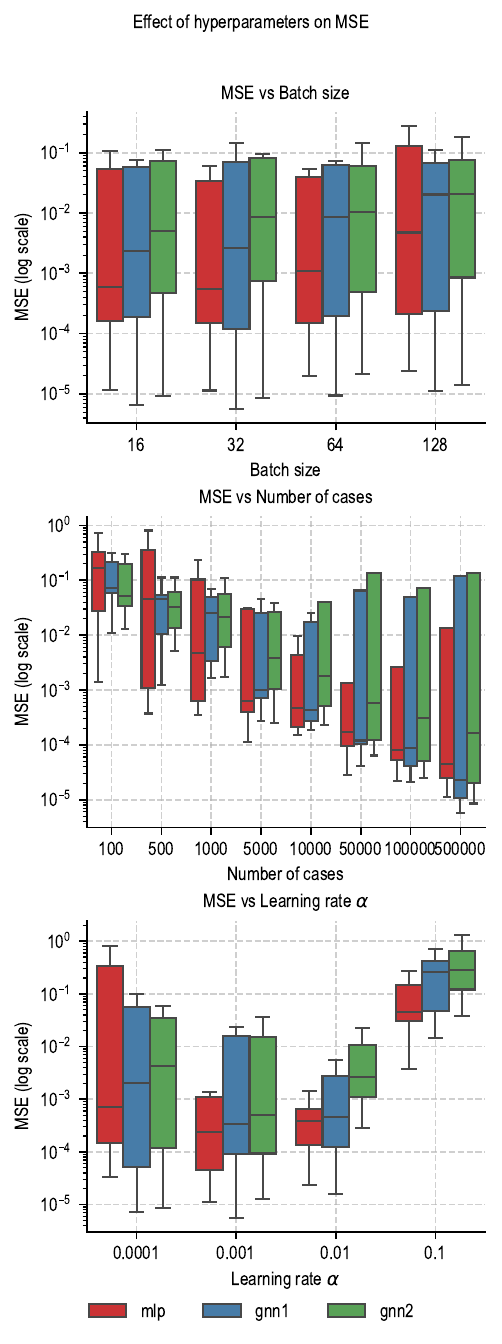}
 \caption{Boxplots of MSE across different hyperparameter configurations.}
 \label{fig:boxplots}
\end{figure}
The results shown in ~\autoref{fig:boxplots} indicate that the training dataset size had the largest and most consistent effect on prediction accuracy across all models.\\
As the number of training samples increased, the test loss decreased notably and the variance across runs was reduced. \\
Batch size had negligible influence on model performance. While no consistent trend for learning rate was observed for GNN1 and the MLP, GNN2 exhibited a strong correlation between learning rate and test loss, suggesting that smaller learning rates improved its stability and final accuracy.\\
\autoref{fig:inftime} shows the inference time\footnote{Windows 11 machine with an Intel Xeon Gold 6226 CPU (16 cores), 10.4 GB RAM, Python 3.9 \& without GPU support} of the considered models in comparison to an N-R solver\footnote{parallelised implementation, one case was solved repeatedly}. 
\\
The lowest computation times are obtained by the MLP, which remains below 0.35 s for 10,000 samples due to its simple architecture. The GNN variants require about 34.6 s (GNN1) and 38.3 s (GNN2) for the same number of samples, while the N-R solver requires about 142.5 s. This corresponds to a speedup of about four for the GNNs and more than four hundred for the MLP compared to N-R. For small sample sizes the advantage of the neural models is less pronounced, as the fixed overhead of the GNNs dominates, whereas for larger case studies the advantage becomes much more significant.
\\
\\
\section{Outlook}
In this paper, sample efficiency in small-scale systems with fixed topology was analyzed using three neural network architectures. It was shown that larger training datasets resulted in lower MSE, an effect expected to become even more critical in larger power grids, as dataset generation becomes increasingly time-consuming. Future work will therefore focus on the integration of physics-informed loss functions, as in \cite{bottcher2023solving}, which explicitly incorporate domain knowledge through known physical relationships, as well as on the application of Known Operator Learning. These approaches are expected to mitigate the dependence on large training datasets.\\
According to \cite{donon2019graph}, GNNs are particularly promising in this regard, as they enable deployment in larger power grids without extensive retraining and reduce the need for large training datasets. The analysis of this capability and its applicability to larger grids will be part of future work.
\section*{Acknowledgment}
This work was conducted within the scope of the research project \textit{GridAssist} and was supported through the “OptiNetD” funding initiative by the German Federal Ministry for Economic Affairs and Energy (BMWE) as part of the 8\textsuperscript{th} Energy Research Programme.

\end{document}